\documentclass[12pt]{article}
\usepackage{amsmath}
\usepackage{amscd}
\usepackage{amssymb}
\usepackage{array}

\begin{document}
\rightline{CERN-PH-TH/2004-097}

\rightline{IFIC/04-28}

\rightline{FTUV-04/0601}

\rightline{UCLA/04/TEP/23}

\newcommand{\R}{\mathbb{R}}
\newcommand{\C}{\mathbb{C}}
\newcommand{\Q}{\mathbb{Q}}
\newcommand{\Z}{\mathbb{Z}}
\newcommand{\Hb}{\mathbb{H}}

\newcommand{\Ss}{Scherk--Schwarz }
\newcommand{\KK}{ Kaluza--Klein }

\newcommand{\cM}{\mathcal{M}}
\newcommand{\cV}{\mathcal{V}}
\newcommand{\cD}{\mathcal{D}}
\newcommand{\cC}{\mathcal{C}}
\newcommand{\cS}{\mathcal{S}}
\newcommand{\cU}{\mathcal{U}}

\newcommand{\rE}{\mathrm{E}}
\newcommand{\ii}{\mathrm{i}}
\newcommand{\rSp}{\mathrm{Sp}}
\newcommand{\rSO}{\mathrm{SO}}
\newcommand{\rSL}{\mathrm{SL}}
\newcommand{\rSU}{\mathrm{SU}}
\newcommand{\rUSp}{\mathrm{USp}}
\newcommand{\rSpin}{\mathrm{Spin}}
\newcommand{\rU}{\mathrm{U}}
\newcommand{\rF}{\mathrm{F}}
\newcommand{\rGL}{\mathrm{GL}}
\newcommand{\rG}{\mathrm{G}}
\newcommand{\rK}{\mathrm{K}}

\newcommand{\fgl}{\mathfrak{gl}}
\newcommand{\fu}{\mathfrak{u}}
\newcommand{\fsl}{\mathfrak{sl}}
\newcommand{\fsp}{\mathfrak{sp}}
\newcommand{\fusp}{\mathfrak{usp}}
\newcommand{\fsu}{\mathfrak{su}}
\newcommand{\fp}{\mathfrak{p}}
\newcommand{\fso}{\mathfrak{so}}
\newcommand{\fl}{\mathfrak{l}}
\newcommand{\fg}{\mathfrak{g}}
\newcommand{\fr}{\mathfrak{r}}
\newcommand{\fe}{\mathfrak{e}}
\newcommand{\ft}{\mathfrak{t}}
\newcommand{\id}{\relax{\rm 1\kern-.35em 1}}

\vskip 1cm

  \centerline{\LARGE \bf  No-scale $D=5$  supergravity from}

 \bigskip

    \centerline{\LARGE \bf  \Ss reduction}

 \bigskip

    \centerline{\LARGE \bf   of $D=6$  theories}

\vskip 1.5cm

\centerline{L. Andrianopoli$^{\flat}$,  S.
Ferrara$^{\flat,\sharp}$ and M. A. Lled\'o$^{\natural}$.}
 \vskip
1.5cm

\centerline{\it $^\flat$ Department of Physics, Theory Division}
\centerline{\it
 CERN, CH 1211 Geneva 23, Switzerland.} \centerline{{\footnotesize e-mail:
Laura.Andrianopoli@cern.ch, \; Sergio.Ferrara@cern.ch}}

\medskip

\centerline{\it \it $^\sharp$INFN, Laboratori Nazionali di
Frascati, Italy.}

\medskip
\centerline{\it $^\natural$
 Departament de F\'{\i}sica Te\`orica,
Universitat de Val\`encia and IFIC}
 \centerline{\small\it C/Dr.
Moliner, 50, E-46100 Burjassot (Val\`encia), Spain.}
 \centerline{{\footnotesize e-mail: Maria.Lledo@ific.uv.es}}

%%%%%%%%%%%%%%%%%%%%%%%%%%%%%%%%%%%%%%%%%%%%%%%%%%%%%%%%%%%%%%%%%%%%%%
%%%%%%%%%%%%%%%%%%%% abstract %%%%%%%%%%%%%%%%%%%%%%%%%%%%%%%%%%%%%%%%
%%%%%%%%%%%%%%%%%%%%%%%%%%%%%%%%%%%%%%%%%%%%%%%%%%%%%%%%%%%%%%%%%%%%%%

\vskip 1.5cm

\begin{abstract}
We perform a generalized dimensional reduction of six dimensional
supergravity theories to five dimensions. We consider the minimal
$(2,0)$ and the maximal $(4,4)$ theories. In each case the
reduction allows us to obtain gauged supergravities of no-scale
type in dimension five with gauge groups that escape previous
classifications. In the minimal case, the geometric data of the
reduced theory correspond to particular cases of the $D=5$ real
special geometry. In the maximal case we find a four parameter
solution which allows partial breaking of supersymmetry.

\end{abstract}

 \vfill\eject

%%%%%%%%%%%%%%%%%%%%%%%%%%%%%%%%%%%%%%%%%%%%%%%%%%%%%%%%%%%%%%%%%%%%%%
%%%%%%%%%%%%%%%%%%%% introduction %%%%%%%%%%%%%%%%%%%%%%%%%%%%%%%%%%%%
%%%%%%%%%%%%%%%%%%%%%%%%%%%%%%%%%%%%%%%%%%%%%%%%%%%%%%%%%%%%%%%%%%%%%%

\section{Introduction \label{introduction}}

In the present paper we discuss the \Ss (SS) dimensional reduction
\cite{ss} on $S^1$ of $D=6$ ungauged supergravity theories with 8 and 32
supercharges. The reduction gives supergravities in five
dimensions with a flat gauge group. Such flat gaugings appear in
four dimensions in the context of both, SS
\cite{adfl1,adfl2,dh,dst} and flux compactifications \cite{fr}.

The \Ss mechanism relies on the presence of a global symmetry
group of the higher dimensional theory. The class of no-scale
supergravities at $D=5$ that we obtain depend then  on the global
symmetry of the $D=6$ theory \cite{ro}.

In the $(2,0)$ (minimal) theories \cite{ro,ns,sa}  there are three
kinds of matter multiplets: the vector multiplet which has no
scalars, the tensor multiplet with 1 scalar and the hypermultiplet
with four scalars.  If we have $n_T$ tensor multiplets and $n_H$
hypermultiplets,  the scalar manifold is a product
\begin{equation}\frac{\rSO(1,n_T)}{\rSO(n_T)}\times
\cM_Q,\label{primo}\end{equation} where $\cM_Q$ is a quaternionic
manifold of quaternionic dimension $n_H$ \cite{bw}.
 The SS phase is, in general, a combination of isometries of
both manifolds.

The graviton multiplet contains a self dual tensor field, while
the tensors from the tensor multiplets are anti-self dual. We
denote the set of tensor fields as $B^r$, $r=0,\dots n_T$, with
$B^0$ pertaining to the graviton multiplet.

 When vector multiplets are present, the vectors
($A^x$, $x=1,\dots n_V$) couple to the tensor fields  and their
interaction term is of the form \cite{fms,frs,rs,ns2}
$$C_{rxy}B^r\wedge F^x\wedge F^y, \qquad F^x=dA^x,$$  with
$C_{rxy}$=constant. This term is related by supersymmetry to the
kinetic term of the vectors
\begin{equation}
C_{rxy}b^rF^x\wedge ^*\!\!F^y.
\label{kv}
\end{equation}
The fields $b^r$, $r=0,\dots n_T$ satisfy the constraint
$$\eta_{rs}b^rb^s=1,$$
which defines the manifold ${\rSO(1,n_T)}/{\rSO(n_T)}$. The
terms (\ref{kv}) explicitly break the $\rSO(1,n_T)$ symmetry, unless the
vector fields $A^x$ transform under some $n_V$-dimensional
representation $R_V$ of $\rSO(1,n_T)$ with the property that
$\mathrm{Sym}(R_V\otimes R_V)$ contains the vector representation.
In that case, the constants $C_{rxy}$ can be chosen as
invariant couplings. This happens, for instance, if $R_V$ is a
spinor representation of $\rSO(1,n_T)$. Remarkably, this choice
leads after  dimensional reduction on  $S^1$ to the real special
geometries which are homogeneous (in particular, symmetric) spaces
\cite{al,ce,dv,dvv}.

Under this assumption, the SS reduction produces a theory with a
flat gauge group of the form $\rU(1)\ltimes R_V$, where the
$\rU(1)$ generator  is in the Cartan subalgebra (CSA) of the
maximal compact subgroup $\rSO(n_T)$ of the global symmetry
$\rSO(1,n_T)$. The $\rU(1)$ group is gauged by the vector coming
from the metric in dimension six. The tensors are in a vector
representation of $\rSO(1,n_T)$, so they are charged under U(1)
(except for some singlets as $B^0$).

We remark that in order to introduce a SS phase in the
tensor-vector multiplet sector it is actually sufficient that the
constants $C_{rxy}$ preserve a $U(1)$ subgroup of $\rSO(1,n_T)$,
which is a much weaker assumption. The examples that we will
consider in this paper have the full $\rSO(1,n_T)$ symmetry.

The generator of the group U(1) may also have a component on the
isometries of the quaternionic manifold \cite{afl}; in particular,
it may have a component in the CSA of the SU(2) R-symmetry, then
breaking supersymmetry (notice that this can happen even if
hypermultiplets are not present, corresponding to a $D=5$
Fayet-Iliopoulos term). The SS reduction leads to a positive
semidefinite potential  also in this case. The $D=5$
interpretation of the theory must correspond to a gauging with the
term $V_R=0$ (see section \ref{scalarpotential}  and Ref.
\cite{gst}).

In the case of (maximal) (4,4) $D=6$ supergravity \cite{sse1}, the
sigma model is
$$\frac GH =\frac {\rSO(5,5)}{\rSO(5)\times \rSO(5)}.$$
The SS phase lies in the CSA of $\rUSp(4)\times\rUSp(4)$
($\rUSp(4)=\rSpin(5)$) so that the theory contains
$4=\mathrm{rank}(\rUSp(4)\times\rUSp(4))$ mass parameters. There
are 16 vectors in six dimensions corresponding to the chiral
(real) spinor representation of $\rSpin(5,5)$, so the flat group
is $\rU(1)\ltimes \R^{16}$ \cite{dst2}. This is a straightforward
generalization of the $D=4$, $N=8$ case studied in Ref.
\cite{adfl1}. Partial supersymmetry breaking to $N=6,4,2,0$ in
$D=5$ may occur depending on how many mass parameters are taken
different from zero.

%%%%%%%%%%%%%%%%%%%%%%%%%%%%%%%%%%%%%%%%%%%%%%%%%%%%%%%%%%%%%%%%%%%%%%
%%%%%%%%%%%%%%%%%%%% Reduction of the (2,0) theory%%%%%%%%%%%%%%%%%%%%
%%%%%%%%%%%%%%%%%%%%%%%%%%%%%%%%%%%%%%%%%%%%%%%%%%%%%%%%%%%%%%%%%%%%%%

\section{Reduction of the (2,0) theory}

We give here the qualitative features of the SS reduction of a
general $(2,0)$ theory from $D=6$ to $D=5$ and show how it
produces an $N=2$ theory in $D=5$ with tensor, vector and hyper
multiplets, and a flat gauge group.

 Let us consider  a $D=6$ theory with $n_T$ tensor multiplets,
 $n_V$ vector multiplets and $n_H$ hypermultiplets. These theories are
 anomalous unless the condition
\begin{equation}n_H - n_V + 29~ n_T =273\label{anomaly}\end{equation} is satisfied \cite{rsss}.

It was shown in Ref. \cite{fms} that when performing a standard
dimensional reduction to $D=5$ on an anomaly-free $(2,0)$ theory,
we obtain a particular class of $N=2$, $D=5$ theories.  After the
reduction, the geometry of the hypermultiplets ($\cM_Q$) remains
unchanged. The scalar manifold of the vector and tensor multiplets
has a real special geometry \cite{gst2}. Let $\cM_R$ be this
manifold in $D=5$ and $d=\dim\cM_R$.

On general grounds, real-special geometry
consists essentially on an embedding of $\cM$ in a manifold of
dimension $d+1$ through a cubic polynomial constraint
$$\mathcal{V} =d_{IJK} t^It^J t^K=1,\qquad I,J,K=1,\dots d+1.$$
The metric induced by the embedding from the metric in the higher
dimensional manifold $a_{IJ}$,
\begin{equation}a_{IJ}=-\frac12\partial_I\partial_J\ln \cV, \qquad g_{ij} =
a_{IJ}\partial_it^I\partial_jt^J|_{\cV=1}, \quad i,j=1,\dots
d.\label{metric}\end{equation} In the following, we will denote
$G_{IJ}=a_{IJ}|_{\cV=1}$. When the $D=5$ theory comes from a
dimensional reduction from $D=6$, $d=n_T+n_V+1$ (the extra scalar
coming from the metric), and the cubic polynomial takes the
particular form
\begin{equation}
\mathcal{V} = 3\left( z\eta_{rs} b^r b^s + C_{rxy}~ b^r a^x
a^y\right) ~; \quad  r=0,1,\cdots n_T ~; \quad x=1,\cdots n_V.
\label{cubic}
\end{equation}
$\eta_{rs}$ is the $(1,n_T)$ Lorentzian metric related to the
space $\rSO(1,n_T)/\rSO(n_T)$ (parametrized by $b^r$) in
(\ref{primo}), $z=\sqrt{g_{55}}=e^\sigma $ is the \KK scalar and $a^x=A^x_6$
are the axions.

We now focus on the cases when  $\rSO(1,n_T)$ is a global
symmetry. This demands the coupling $C_{rxy}$ to be an invariant
coupling in the sense explained in Section \ref{introduction}. One
could then introduce a SS phase in the CSA of $\rSO(n_T)$. Some of
the vector and tensor multiplets are charged under this generator,
so they acquire mass. In the $D=5$ interpretation the vectors
gauge a non-abelian
 flat group, but their scalar partners give no  contribution to the scalar
potential, in agreement with the known results on $D=5$ gauged
supergravity \cite{gz,cd}. The gauging of flat groups in the
context of $N=2$ supergravity has not been considered in previous
classifications \cite{egz}. These gaugings are always of no-scale
type due to the particular structure of the critical points
\cite{cfkn}.

Finally, we want to note that to uplift (oxidate \cite{cjlp,ke})
to $D=6$ a five dimensional $N=2$ supergravity a necessary
condition is that the cubic polynomial defining the real special
geometry has the form (\ref{cubic}). All the homogeneous spaces
with real special geometry fall in this category. These spaces
have been classified in Refs. \cite{al,ce,dv,dvv}; they were
denoted as $L(q,P,\dot P)$ in Ref. \cite{dvv}. We explain here
this notation. Let $q=n_T-1$ and let $\cD_{n_T}$ be the real
dimension of an irreducible representation of $\rSpin(1,n_T)$. For
$n_T=1,5$ mod 8 there are two inequivalent real or pseudoreal
(quaternionic) representations. Let  $P$ and $\dot P$ denote the
number of copies of such representations ($\dot P=0$ for $n_T\neq
1,5$ mod 8). Then, $n_V= (P+\dot P)\cD_{n_T}$. The R-symmetry
group of $\rSpin(1,n_T)$ in the representation $(P, \dot P)$ is
denoted by $\cS_q(P,\dot P)$ (see Table 3 of Ref. \cite{dvv}).

When $\dot P=0$, the notation $L(q,P)=L(q, P,0)$ is used. The
symmetric spaces \cite{gst2} correspond to the particular cases
$L(1,1)$, $L(2,1)$, $L(4,1)$, $L(8,1)$, $L(-1,P)$ and $L(0,P)$. We
also have $L(q,0)= L(0,q)$. They are reported in Table 2. of Ref.
\cite{dvv}. The examples of SS reductions reported in this paper
will actually fall in this class.

%%%%%%%%%%%%%%%%%%%%%%%%%%%%%%%%%%%%%%%%%%%%%%%%%%%%%%%%%%%%%%%%%%%%%%
%%%%%%%%%%%%%%%%%%%% Tensor multiplet sector%%%%%%%%%%%%%%%%%%%%%%%%%%
%%%%%%%%%%%%%%%%%%%%%%%%%%%%%%%%%%%%%%%%%%%%%%%%%%%%%%%%%%%%%%%%%%%%%%

\subsection{Tensor multiplet sector}

The $D=5$ theory obtained through an ordinary \KK dimensional
reduction contains $n_T+n_V+1$ vector multiplets. This is  because
the  (anti) self-duality condition in $D=6$
\begin{equation}
\partial_{[\mu} B^r_{\nu\rho]}=\pm \frac 1{3!}\epsilon_{\mu\nu\rho\lambda\tau\sigma} \partial^\lambda B^{r|\tau\sigma}
 \qquad \mu,\nu =1,\dots 6\label{selfdual6}
\end{equation}
tells us that in $D=5$ the two form $B_{\mu\nu}^r$ is dual to the
vector $B_{\mu 6}^r$ ($\mu,\nu =1,\dots 5$).

We want  now to perform
   a SS generalized dimensional reduction instead. Let $M^{r}_{\ s} = -M_{s}^{\
   r}$ be the SS phase
 in the CSA of the global symmetry $\rSO(n_T)\subset \rSO(1,n_T)$.
 The form $B^0$ (of the gravitymultiplet) is inert under
 $\rSO(n_T)$, so in the rest of this subsection the value $r=0$ is
 excluded and $r=1,\dots n_T$.
The $D=6$ anti self-duality condition gives now
\begin{eqnarray}
\partial_{[\mu} B^{r}_{\nu\rho]}&=&\frac 1{3!}\epsilon_{\mu\nu\rho\lambda\tau 6}
\left(\partial^6 B^{r|\lambda\tau} +2\partial^\lambda B^{r|\tau
6}\right) =\nonumber \\&&\frac
1{3!}\epsilon_{\mu\nu\rho\lambda\tau }\left(M^{r}_{\ s}
B^{s|\lambda\tau} + F^{r |\lambda\tau}\right),\qquad  \mu,\nu
=1,\dots 5. \label{selfdual5}
\end{eqnarray}
where $F^{r}_{\lambda\tau}= 2\partial_{[\lambda} B^{r}_{\tau ]6}$.

Equation (\ref{selfdual5}) can be rewritten as a self-duality
condition for a massive two-form in five dimensions \cite{tpv}.
Assume that the Cartan element $M$ is invertible; then we can
define
$$\hat B^{r}_{\mu\nu}= B^{r}_{\mu\nu} + (M^{-1})^{r}_{\ s} F^{s}_{\mu\nu},$$
so
$$\partial_{[\mu} \hat B^{r}_{\nu\rho]}=M^{r}_{\
s}\frac 1{3!}\epsilon_{\mu\nu\rho\lambda\tau}\hat
B^{s|\lambda\tau},$$ that is, $$d\hat B^{r}={M^{r}_{\ s}} ^*\!\hat
B^{s}.$$ For $n_T$ even, an element  $M$  with non zero
eigenvalues
 $\pm im_\ell\neq 0$  ($\ell =1,\dots n_T/2$) is invertible.  Then we have
$n_T/2$ complex massive two-forms. For $n_T$ odd, the matrix $M$
has at least one zero-eigenvalue. The corresponding antisymmetric
tensor $B^{r_0}$
 is a gauge potential which can be dualized to a vector. If some
 other eigenvalue $m_\ell$ is zero, the same argument applies and
 there will be a couple of
 tensors (or one complex tensor) which can be dualized to vectors.

Summarizing, in the five dimensional theory there are  $2n\leq
n_T$,  massive tensor multiplets (or $n$ complex ones) and
$n_T-2n+1$ abelian vector multiplets, one of them formed with  the
vector which is dual (after reduction to $D=5$) to the self dual
tensor present in the $D=6$ graviton multiplet. This vector is a
singlet of the global symmetry group.

%%%%%%%%%%%%%%%%%%%%%%%%%%%%%%%%%%%%%%%%%%%%%%%%%%%%%%%%%%%%%%%%%%%%%%
%%%%%%%%%%%%%%%%%%%% The scalar potential%%%%%%%%%%%%%%%%%%%%%%%%%%%%%
%%%%%%%%%%%%%%%%%%%%%%%%%%%%%%%%%%%%%%%%%%%%%%%%%%%%%%%%%%%%%%%%%%%%%%

\subsection{The scalar potential in $D=5$\label{scalarpotential}} In this section we
compute the scalar potential of the SS reduced theory.

The scalar potential comes from the kinetic term of the scalar
fields \cite{ss}. The only scalars at $D=6$ are in the tensor and
hyper multiplets, which parametrize the manifold in (\ref{primo}).
We denote by $\varphi^i$, $i=1,\dots n_T$ the coordinates on
$\rSO(1,n_T)/\rSO(n_T)$, and let
$$v^a=v^a_i\partial_\mu\varphi^i dx^\mu=v^a_\mu dx^\mu, \qquad a=1,\dots n_T$$
be the pull back to space time of the vielbein one form.
Similarly, the quaternionic manifold $\cM_Q$ \cite{bw,abcdffm},
with holonomy $\rSU(2)\times \rUSp(2n_H)$, has coordinates $q^u$,
$u=1,\dots 4n_H$ and vielbein
$$\cU^{\alpha A}=\cU^{\alpha A}_u\partial_\mu q^u dx^\mu=\cU^{\alpha A}_\mu
dx^\mu, \qquad \alpha=1,\cdots 2n_H,\quad A=1,2.$$ There is still
a scalar mode coming from the metric $e^\sigma =\sqrt{ g_{55}}$.

For the scalar potential we obtain
\begin{equation}
V(\sigma , \varphi , q) =V^{SS}_T+V^{SS}_H=  e^{-\frac 83 \sigma}
\left[v_6^a(\varphi) v_{6 a}(\varphi) + {\mathcal {U}}_6^{\alpha
A}(q)\, {\mathcal {U}}_6^{\beta B}(q) \C_{\alpha\beta}
\epsilon_{AB}\right], \label{5dpot}
\end{equation}
where $\C_{\alpha\beta}$ and $\epsilon_{AB}$ are the antisymmetric
metrics.

We see that this potential is semipositive definite. The critical
points occur at \begin{equation} v^a_6 (\varphi )=0 \qquad \mbox{
and } \qquad {\mathcal {U}}_6^{\alpha
A}(q)=0,\label{critical}\end{equation} so $V=0$ at the critical
points, which are then Minkowski vacua. The scalar $\sigma$ is not
fixed, so the theory is of no-scale type. Notice that
(\ref{critical}) implies
$$ v^a_6 (\varphi )=v_i^aM_i^j\varphi^j=0, \qquad \cU_6^{\alpha
A}(q)=\cU_u^{\alpha A}M_v^uq^{v}=0.$$
 If
the mass matrices have some vanishing eigenvalues, then this
results in some moduli of the theory, other than $\sigma$. For
$n_T$ odd, since the tensor multiplet mass matrix has always one
vanishing eigenvalue, there are at least two massless scalars.
There are three  massless vectors  in this case.

The SS potential given in (\ref{5dpot}) should be compared to the
most general gauging of $N=2$, $D=5$ supergravity
\cite{gz,cd,bcdgvv}
$$V_{D=5}= V_T + V_H + V_R ~; \qquad V_T \geq 0 ~; \quad V_H \geq 0,$$
where $V_T$ and $V_H$   are the contributions of tensor and
hypermultiplets (separately positive) and $V_R$ is the
contribution from vector and  gravity multiplets due to the
quaternionic Killing prepotential $P^X_I$, $X=1,2,3$
\cite{abcdffm}.

For a $D=5$ gauging corresponding to a SS reduction, we then need $V_R=0$.

The explicit form of $V_R$ is \cite{cd,gz,bcdgvv}
\begin{equation}V_R = -4 t_{IJ} G^{IK} G^{JL}P^X_KP^X_L = -\frac 43
\left(\frac 13 (t^{-1})^{IJ} + t^I t^J\right)
P^X_IP^X_J\label{killingprepotential}\end{equation} where $t_{IJ}
=d_{IJK} t^K$ and $G^{IJ} = a_{IJ}|_{\mathcal{V}=1}= -\frac 13 (t^{-1})^{IJ} + t^I t^J$
(see (\ref{metric})).

Even when there are no hypermultiplets, this term is not
necessarily zero, because one can take a constant prepotential,
$P^{X_0}_I=g_I=$constant (the rest zero.). $g_I$ is the $N=2$
Fayet-Iliopoulos parameter, and we retrieve the particular form of
$V_R$ found in Ref. \cite{gst}.

Equation (\ref{killingprepotential}) can also be written as
\begin{equation}V_R = -4d^{IJK}t_IP^X_JP^X_K,\label{fayetili}\end{equation} where indices are lowered and
raised with the metric $G_{IJ}$. For symmetric spaces one has
$d_{IJK}=d^{IJK}$ \cite{gst}.

From the point of view of the SS reduction, the constant
prepotential corresponds to an SU(2) phase, which in absence of
hypermultiplets only gives masses to the fermions. Therefore we
must have $V_R=0$ for any value of $t^I$ in the reduced theory.
Moreover, since this depends only on the real special geometry
(see (\ref{fayetili})), this conclusion also holds in presence of
hypermultiplets.

In the SS reduction the vector gauging the $\rU(1)\subset \rSU(2)$
is the partner of the scalar $z=e^\sigma$, so $P^X_z\neq 0$ and
the rest are zero. $V_R=0$ then requires
\begin{equation}(t^{-1})^{zz} = -3
(t^z)^2\quad \Longleftrightarrow \quad
G^{zz}=2(t^z)^2.\label{inverse}\end{equation}

Let us  consider some  particular examples of theories with
$V_R=0$. Setting $n_V=0$, equation (\ref{cubic}) becomes
$$\cV=3z\eta_{rs}b^rb^s$$
and one can check that (\ref{inverse}) holds \cite{gst}. It also
holds for the  spaces $L(0,P)$. More generally, it holds for all
symmetric spaces with real special geometry because of the relations
$d_{IJK}=d^{IJK}$ and   $d_{zzI}=0$. They readily imply $V_R=0$.

 We have checked that there are in fact counterexamples
to the condition $V_R=0$ among the theories classified in
\cite{dv,dvv} which are all of the form (\ref{cubic}),  so $V_R=0$
is a further restriction satisfied by the $D=5$ real geometries
that can be uplifted (oxidated) to $D=6$.  It would be interesting
to know, in the general case, what  the conditions on the
coefficients $C^r_{xy}=\eta^{rs}C_{sxy}$ in (\ref{cubic}) are to
have $V_R=0$.

We will see in the next subsection that the possible resolution of
this puzzle lies in the cancellation of anomalies of the six
dimensional theory.

%%%%%%%%%%%%%%%%%%%%%%%%%%%%%%%%%%%%%%%%%%%%%%%%%%%%%%%%%%%%%%%%%%%%%%
%%%%%% Conditions for uplifting  $D=5$ to $D=6$ theories %%%%%%%%%%%%%
%%%%%%%%%%%%%%%%%%%%%%%%%%%%%%%%%%%%%%%%%%%%%%%%%%%%%%%%%%%%%%%%%%%%%%

\subsection{Conditions for uplifting  $D=5$ to $D=6$ theories}

In $D=6$, $(2,0)$ chiral theories it was found that there is, in
general, a clash between the gauge invariance of the  two-forms
and the gauge invariance of the 1 forms (vector fields). For
generic couplings $C_{rxy}$, in the abelian case, the $n_V$ U(1)
currents $J_x$ are not conserved but satisfy the equation \cite{ns2,fms}
\begin{equation}d^*\!J_x=\eta_{rs}C^r_{xy}C^s_{zw}F^y\wedge F^z\wedge F^w.\label{notconserved}\end{equation} This
violation of the gauge invariance implies also a violation of
supersymmetry because the theory is formulated in the Wess--Zumino
gauge and the supersymmetry algebra closes only up to gauge
transformations \cite{fms}. The current is conserved if the
constants $C^r_{xy}$ satisfy the condition
\begin{equation}\eta_{rs}C^r_{x(y}C^s_{zw)}=0.\label{symmetry1}\end{equation}
This condition is equivalent
to the seemingly stronger condition
\begin{equation}\eta_{rs}C^r_{(xy}C^s_{zw)}=0\label{symmetry2}\end{equation}
because $C^r_{xy}=C^r_{yx}$. This can also
be seen from the fact that the anomaly polynomial \cite{fms}
$$A\sim \eta_{rs}C^r_{xy}C^s_{zw}F^x\wedge F^y\wedge F^z\wedge F^w$$
vanishes if (\ref{symmetry2}) holds. It is interesting to observe
in this respect that among the homogeneous spaces in Ref.
\cite{dv,dvv} only the symmetric spaces, with the exception of the
family $L(-1,P)$, $P>0$, satisfy this condition \cite{dv,dvv}.

Also, we must note that the symmetric spaces  satisfying
(\ref{symmetry2}) do have in fact $V_R=0$, while for the
homogeneous, non symmetric cases there are counterexamples.

Condition (\ref{symmetry1}) is only required for a $D=6$ ungauged
supergravity. If the theory in $D=6$ is already gauged, the terms
in the right hand side of (\ref{notconserved})
 may be compensated by (one loop) quantum
anomalies through a Green-Schwarz mechanism, namely, the Lagrangian
becomes a Wess-Zumino term \cite{sa,frs,ns2}. The $D=6$ potential
 is semipositive definite and simply given by \cite{ns}
$$V_{D=6}\simeq \sqrt{g}P_x^XP_y^X(C^{-1})^{xy}, \qquad \hbox{where }
C_{xy}=C_{rxy}b^r.$$ The $D=6$ supersymmetric vacua occur at
$P_x^X=0$. An hypermultiplet can be ``eaten" by a vector
multiplet, making it massive. Note that there are not BPS particle
multiplets in $D=6$. The additional contribution to the potential
in $D=5$ is
$$\sqrt{g_5}\,e^{-\frac 23\sigma}P_x^XP_y^X(C^{-1})^{xy}.$$ Since in this case $V_R$ needs not to
vanish, one may find new vacua in the SS reduction.

As an illustration of spaces satisfying (\ref{symmetry2}), we give
the spectrum of tensor, vector and hypermultiplets  for the
exceptional symmetric spaces in Table \ref{exceptional}.
\begin{table}[ht]
\begin{center}
\begin{tabular} {|c ||c| c|c|c|}
\hline  $L(q,P)$&$L(1,1)$ & $L(2,1)$ &$L(4,1)$& $L(8,1)$\\
\hline
$(n_T,n_V,n_H)$ &(2,2,217) & (3,4,190) &(5,8,136)& (9,16,28) \\
\hline
\end{tabular}
\caption{Exceptional symmetric spaces}\label{exceptional}
\end{center}
\end{table}
Note that the values of $n_T$ and $n_V$ are given by the uplifting
(oxidation) procedure of Ref. \cite{ju,ke}. These spaces are
contained in the classification of Refs. \cite{dv,dvv} and
consequently have a cubic polynomial of the form (\ref{cubic}).
The number $n_H$ instead is fixed by the gravitational anomaly
cancellation (\ref{anomaly}). For generic SS phases, the $L(1,1)$
model  has one massless scalar and two massless vectors. All the
other exceptional models have two massless scalars and three
massless vectors.

There are no other solutions in the series $L(q,P,\dot P)$. It is
obvious that for non homogeneous spaces the constants $C^r_{xy}$
are rather arbitrary and there may be much more solutions to the
uplifting condition.

However, in order to have a SS phase in the tensor and vector
multiplet sector, non homogeneous spaces should have at least a
residual U(1) isometry.

We therefore find that a possible explanation to the fact that
$V_R\neq 0$ in  $D=5$ supergravity with cubic form (\ref{cubic})
may be connected to the violation of supersymmetry in the six
dimensional theory.

%%%%%%%%%%%%%%%%%%%%%%%%%%%%%%%%%%%%%%%%%%%%%%%%%%%%%%%%%%%%%%%%%%%%%%
%%%%%%%%%%%%% SS reduction of the $N=(4,4)$, $D=6$ theory %%%%%%%%%%%%
%%%%%%%%%%%%%%%%%%%%%%%%%%%%%%%%%%%%%%%%%%%%%%%%%%%%%%%%%%%%%%%%%%%%%%

\section{SS reduction of the $N=(4,4)$, $D=6$ theory}

Let us sketch here the general features and mass spectrum of the
$D=5$ theory obtained by SS compactification from the maximal
supergravity in $D=6$.

The gravitational multiplet of the six dimensional theory contains
the graviton $e_\mu^m$, four  gravitini $\psi_A$, $A =1,\dots 4$
in the fundamental of $\rSp(4)_L$, four gravitini $\psi_{\tilde
A}$, $\tilde A =1,\cdots 4$   in the fundamental of $\rSp(4)_R$,
five self dual and five anti-self dual 2-form potentials
$B^r_{+}$, $B^{\dot r}_{-}$,\, $r,\dot r =1,\cdots 5$ in the
fundamental of $\rSO(5,5)$,  16 vector potentials $A^\alpha_\mu$,
 $\alpha =1,\cdots 16$ in the spinorial of $\rSO(5,5)$, 20 dilatini
 $\chi^{r\tilde A}$ in the $({\bf 5},{\bf 4})$ of $\rUSp(4)_L\times \rUSp(4)_R$,
20 dilatini $\chi^{A \dot r}$  in the $({\bf 4},{\bf 5})$ of
$\rSp(4)_L\times \rSp(4)_R$, and 25 scalars $\varphi^{r\dot s}$
spanning the scalar manifold $\rSO(5,5)/(\rSO(5)\times \rSO(5))$.

The  global symmetry of the theory is the maximal compact subgroup
of $\rSpin(5,5)$,  $\rUSp(4)_L\times \rUSp(4)_R$, so that one can
turn on an SS phase in its CSA. Since the rank is 4, we have 4
mass parameters $m_i, \tilde m_\ell$ ($i,\ell =1,2$).

In  $D=5$ we obtain a maximal supergravity  gauged with the flat
group $\rU(1) \ltimes {\bf 16}$. The $\rU(1)$ factor in the CSA is
gauged by the \KK graviphoton $B_\mu$ and the 16 translations are
gauged by the vectors $Z_\mu^\alpha = A_\mu^\alpha - A_6^\alpha
B_\mu$. This is a flat subgroup of $\rE_{6(6)}$, according to the
Lie algebra  decomposition \cite{dst2}
$$\fe_{6(6)} \to \fso(5,5) \oplus  \fso(1,1) + {\bf 16}^+ + {\bf 16}^-,$$
and it gives a  gauging of $N=8$, $D=5$ supergravity not included
in previous classifications \cite{acfg}.

For generic values of $m_i$, $m_\ell$  all the $Z^\alpha$ vector
fields become massive through the Higgs mechanism, with masses $|
m_i \pm \tilde m_\ell|$. This can be understood from the fact that
the spinorial representation  $(\mathbf{16})$ of $\rSO(5,5)$ is
the $(\mathbf{4}, \mathbf{4})$ of $\rUSp(4)_L\times \rUSp(4)_R$,
so that the \Ss phase in this sector is $M^A_{\ B}\otimes
1+1\otimes M^{\tilde A} _{\ \tilde B}$, with eigenvalues  $\pm \ii
m_i \pm \ii \tilde m_\ell $. On the other hand, the \KK
graviphoton $B_\mu$ stays massless. The 16 axions have been
absorbed by the 16 vectors to become massive.

In the scalar sector the SS phase appears in the kinetic terms
through
$$\partial_6\varphi^{r\dot r}=  M^r_{\ s}\varphi^{s \dot r}+
 \tilde M^{\dot r}_{\ \dot s} \varphi^{r \dot s},$$
where the antisymmetric matrices $M^r_{\ s}$, $ \tilde M^{\dot
r}_{\ \dot s}$  have eigenvalues $\pm \ii (m_1\pm m_2)$, 0 and $
\pm \ii(\tilde m_1\pm \tilde m_2)$, 0 respectively. This builds up
the (positive-definite) $D=5$ scalar potential
\begin{equation}
V(\sigma , \varphi ) =  e^{-\frac 83 \sigma}P_6^{r\dot s}(\varphi) P_{6 r\dot s}(\varphi) .
\label{maxpot}
\end{equation}
which vanishes for $P_6^{r\dot s}(\varphi)= 0$.

From  the $D=5$ gauged supergravity point of view this corresponds
to gauge an isometry in $\rE_{6(6)}/\rUSp(8)$, with Killing vector
$$K=k^{r\dot r}\frac {\partial}{\partial \varphi^{r\dot r}}= (M^r_{\ s}\varphi^{s \dot r}+
 \tilde M^{\dot r}_{\
\dot s} \varphi^{r \dot s})\frac {\partial}{\partial
\varphi^{r\dot r}}.$$ We have that 24 out of the 25 scalars become
massive, 16 of them with masses $|\pm m_1 \pm m_2 \pm \tilde m_1
\pm \tilde m_2|$, 4 of them with masses $|\pm m_1 \pm m_2 |$, 4 of
them with masses $| \pm \tilde m_1 \pm \tilde m_2|$, with only one
massless scalar, other than the \KK scalar
$\sqrt{g_{55}}=e^\sigma$.

As far as the antisymmetric tensors $(B^r_+,B^{\dot r}_-)$  are
concerned, the vector representation {\bf 10} of $\rSO(5,5)$
decomposes under the maximal compact subgroup
$$\mathbf{10} \to
 (\mathbf{5},\mathbf{1}) + (\mathbf{1},\mathbf{5}).$$
The SS phase of this sector is
$$M^r_{\ s}\oplus \mathbf{0}_{5\times 5} + \mathbf{0}_{5\times 5} \oplus \tilde M^{\dot r}_{\ \dot s}.$$
Correspondingly, we have in $D=5$ four complex antisymmetric
tensors, two  with masses $|m_1 \pm m_2|$ and two with masses
$|\tilde m_1 \pm \tilde m_2|$, plus 2 massless tensors which may
be dualized to abelian vectors. They do not participate in the
gauging so they stay massless.

All the gravitini $\psi_A, \psi_{\tilde A}$ become massive, with masses (equal in couples)
 $|m_i|, |\tilde m_\ell|$ respectively.

The dilatini $\chi^{r\tilde A}, \chi^{A\dot r}$ also get masses.
 16 of them have masses $|\pm m_1 \pm m_2 \pm \tilde
m_\ell|$, 16 have masses $|\pm m_i \pm \tilde m_1 \pm \tilde m_2
|$, 4 have masses $ |\pm \tilde m_\ell|$ and 4 have masses $|\pm
m_i|$.

We observe that the moduli space of this theory contains two
scalars and locally is $\rSO(1,1) \times \rSO(1,1)$.
If we set to zero one of the mass parameters ({\em e.g.} $m_1=0$)
 we get an unbroken $N=2$, $D=5$ supergravity with two massless vector multiplets.

 There are two ways of getting $N=4$ supersymmetry, either we set
 $m_1=m_2=0$ or $m_1=\tilde m_1=0$. Finally, setting $m_1=m_2=\tilde
 m_1=0$ we get an $N=6$ theory \cite{cr}.

\section*{Acknowledgements}

M. A. Ll. wants to thank the Physics and Mathematics Departments
at UCLA for their hospitality during the realization of this work.

 The work of S.F. has been supported in
part by the D.O.E. grant DE-FG03-91ER40662, Task C, and in part by
the European Community's Human Potential Program under contract
HPRN-CT-2000-00131 Quantum Space-Time, in association with INFN
Frascati National Laboratories.

The work of M. A. Ll. has been supported by the research grant BFM
2002-03681 from the Ministerio de Ciencia y Tecnolog\'{\i}a
(Spain) and from EU FEDER funds and by D.O.E. grant
DE-FG03-91ER40662, Task C.

S. F. wants to thank C. Angelantonj and specially A. Sagnotti for
illuminating discussions.

\end{document}